\documentclass[letter]{aa}

\usepackage{multirow}
\usepackage{graphicx}
\usepackage{natbib,twoopt}
\usepackage[bookmarks=false]{hyperref} 
\usepackage{subfigure}
\usepackage{float}
\usepackage{array}
\usepackage{epstopdf}

\hypersetup{draft}
\begin{document}

\title{A deep X-ray spectral imaging of the bow-shock pulsar wind nebula associated with PSR B1929+10}

\author{Sangin Kim$^{1}$
\and C.Y. Hui$^{1}$\thanks{E-mail: huichungyue@gmail.com}
\and Jongsu Lee$^{1}$
\and Kwangmin Oh$^{1}$
\and L.C.C. Lin$^{2}$
\and J. Takata$^{3}$}

\institute{Department of Astronomy and Space Science, Chungnam National University, Daejeon 34134, Korea 
\and
Department of Physics, UNIST, Ulsan 44919, Korea
\and
Institute of Particle physics and Astronomy, Huazhong University of Science and Technology, China}


\abstract
{ In this work, we study the X-ray bow-shock nebula powered by the mature pulsar PSR B1929+10 using data from  \emph{XMM-Newton,}  with an effective exposure of $\sim$ 300 ks, offering the deepest investigation of this system thus far.
 We found the X-ray axial outflow extends as long as $\sim$ 8 arc minute behind the proper motion direction, which is a factor of two longer than the result reported in the previous study. Furthermore,
 we found evidence of two faint lateral outflows extending laterally with respect to the proper motion.
 We also found indications of spectral hardening along the axial outflow, suggesting that certain acceleration processes might occur along this feature.}


\titlerunning{X-ray spectral imaging of PSR B1929+10's PWN}
\authorrunning{Kim et al.}

\maketitle
\section{Introduction}
 An asymmetric supernova explosion can result in a kick velocity on the stellar remnant.
 The average projected two-dimensional velocity of the pulsar population in our Galaxy is $\sim250$~km s$^{-1}$ (Hobbs et al. 2005) is derived from this effect. 
 When such fast-moving pulsars traverse the interstellar medium (ISM) with supersonic speed, a bow-shock pulsar wind nebula (PWN) may be formed. 

PSR B1929+10 is one of the nearest pulsars to Earth, with a parallax distance of $d=361^{+10}_{-8}$ pc (Chatterjee et al. 2004).
 At this distance, PSR B1929+10 has a projected velocity of $V_{\bot}=177^{+4}_{-5}$ km s$^{-1}$.
 The spin period and the spin-down rate of PSR~B1929+10 are found to be $P=226.5$ ms and $\dot{P}=1.16\times10^{-15}$~s/s, respectively (Manchester et al. 2005).
 These measurements imply that it is a ``middle-aged" mature pulsar with a spin-down age of $\tau \sim 3\times10^{6}$~years and a spin-down power of $\dot{E}=3.9\times10^{33}$ ergs s$^{-1}$. It is one of the oldest known pulsars with an X-ray nebula.
 Its X-ray emission was first discovered with the {\it Einstein} Observatory by Helfand (1983) and its X-ray pulsations were detected in a deep \textit{ROSAT} observation by Yancopolous et al. (1994).

 Despite having a relatively old age and low spin-down power, its proximity makes it appear as one of the brightest mature pulsars associated with a bow-shock PWN (Kargaltsev et al. 2008).
 There have been several investigations performed on PSR~B1929+10's X-ray PWN (Wang et al. 1993; Becker et al. 2006; Hui et al. 2008; Misanovic et al. 2008).
 The extended X-ray emission that trails the proper-motion of PSR~B1929+10 was initially discovered by Wang et al. (1993) using the data acquired by the {\it ROSAT} position-sensitive proportional counter (PSPC).
 A {\it XMM-Newton} follow-up study using the MOS1/2 cameras on board with an effective exposure of $\sim60$~ks confirmed the PWN detection and placed a constraint of $\sim4$~arcmin on its length (Becker et al. 2006).
 Furthermore, Becker et al. (2006) found that its spectrum can be modeled with an absorbed power-law model with a photon index of $\Gamma=2.0\pm0.4$.
 With the sub-arcsecond spatial resolution of {\it Chandra}, Hui et al. (2008) have discovered an asymmetric arc-like structure with an extent of $\sim10$~arcsec at the head of the bow-shock nebula.
 The authors have speculated that the asymmetry can be a result of anisotropic pulsar wind as well as inhomogeneities in the surrounding ISM (Hui et al. 2008). Such compact X-ray nebula is also discovered in the independent analysis of the same data reported by Misanovic et al. (2008).

 Although the aforementioned studies suggest the nonthermal extended X-rays originate from the bow-shock, there are many properties of PSR~B1929+10's PWN are yet to be explored.
 For example, it is still unknown whether there is any spectral variation across the feature which may result from synchrotron cooling or other acceleration processes in the extended feature (e.g., Hui et al. 2017).
 It is also possible that a PWN can have several spatial components, such as in the Crab Nebula (Weisskopf et al. 2000) and Geminga (Hui et al. 2017; Posselt et al. 2017).
 To address these questions, an X-ray spectral imaging analysis with an exposure that is much deeper than in the aforementioned studies is necessary and this need serves as the motivation for our investigation.

\section{Observations and data reduction}
 We  used all the archival data obtained by {\it XMM-Newton} in our investigation.
 There have been seven observations of PSR B1929+10 carried out by the cameras onboard the European Photon Imaging Camera (EPIC) over $\sim$15 years.
 As the PN camera in these observations operated in small window mode, its field does not have full coverage of the extended X-ray feature. 
 Therefore, we only used the data obtained by MOS1 and MOS2 in our work, which  operated in full window mode.
 This  enables a spatial analysis of the diffuse emission in a 30 arcmin field of view.

 We used the {\it XMM-Newton} Science Analysis System (XMMSAS version 16.1.0) for calibration and data reduction. Utilizing the XMMSAS task \texttt{emproc} with the latest  calibrated files available (CCFs, released in 2019 June), we compiled the MOS1/2 event lists.
 We  applied conservative filtering on the calibrated data with \texttt{PATTERN} selected between 0 and 12 and set the \texttt{FLAG} equal to zero (for removing events registered by hot pixels).
 Subsequently, we  removed the time intervals where the backgrounds are high.
 After the good time interval (GTI), the time segments of count rates above the values specified under ``\texttt{RATE} cut" in Table~\ref{tab:Exposures} were filtered and removed, with the effective exposures given the fifth column of Table~\ref{tab:Exposures}..
 We also note that all the data analyses in our work are limited in the energy range of 0.3-10 keV.

\section{Data analysis}
\subsection{Spatial analysis}

\begin{figure}
\centering
        \includegraphics[width=\columnwidth,trim={0.5cm 2.5cm 0.1cm 0cm},clip]{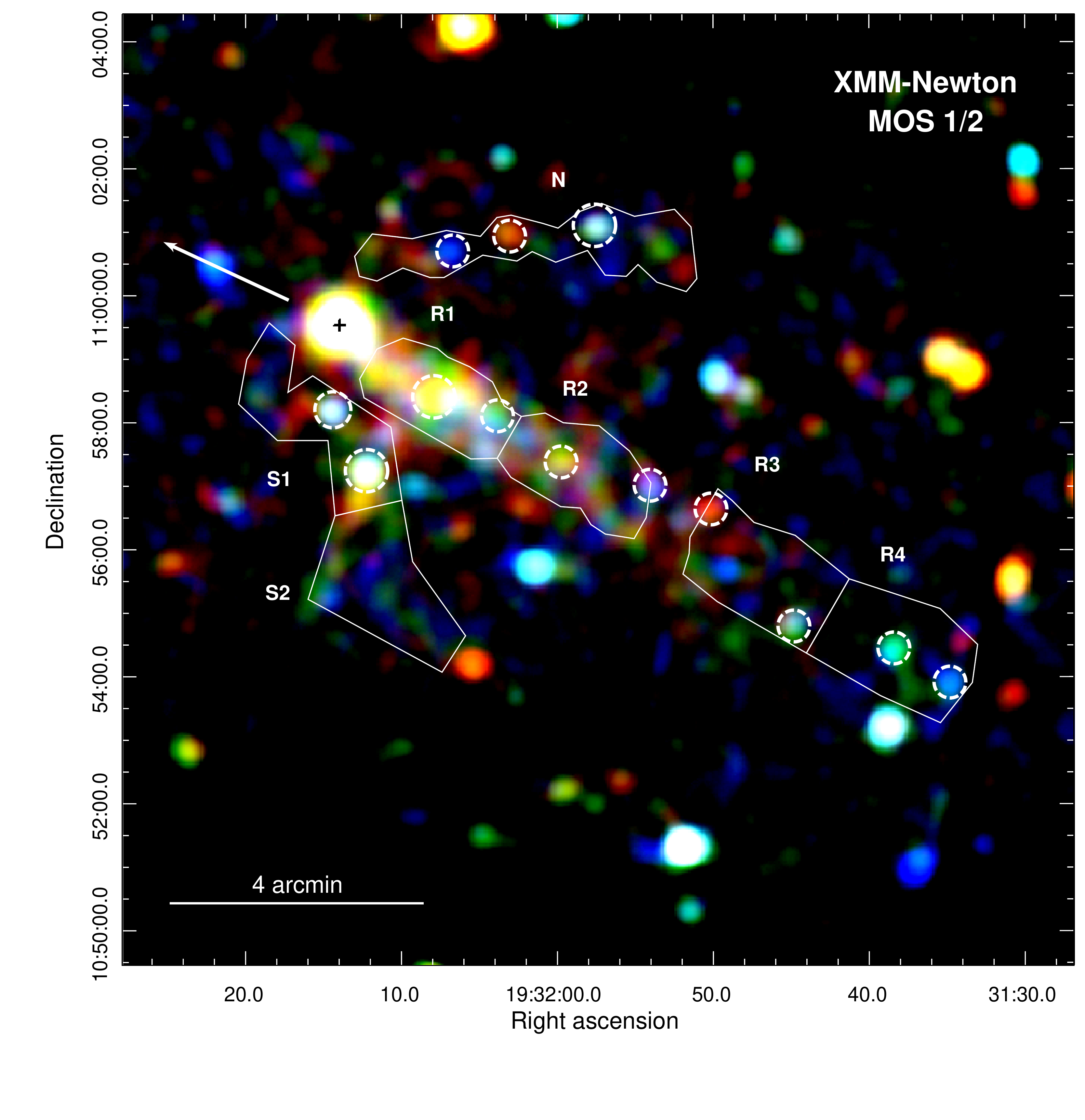}
        \caption{Color-coded MOS1/2 image of the field around PSR B1929+10 with merged vignetting-corrected and background-subtracted data. The pulsar position and its proper motion are illustrated by a cross and an arrow, respectively. The adopted regions for spatially-resolved analysis, R1, R2, R3, R4, S1, S2, and N, are labeled accordingly.}
        \label{fig:images}
\begin{flushleft}
\end{flushleft}
\end{figure}

 Our main goals are focused on constraining the scale of the tail-like emission revealed by Wang et al. (1993; referred to as axial outflow hereafter) and searching for possible additional structure(s) of diffuse emission associated with PSR B1929+10. 
 Before we merged all the data, we corrected for the vignetting effect in the data obtained by each camera in individual observations.
 We generated the vignetting-corrected background-subtracted images for each camera in each individual observation by using the XMMSAS task \texttt{eimageget}.  
 The corrected images were subsequently merged. 
 Figure~\ref{fig:images} shows a merged color-coded MOS1/2 image (red: 0.3-1.0 keV, green: 1.0-2.0 keV, blue: 2.0-10.0 keV) which is smoothed by a Gaussian kernel with a size of $11$~arcsec.

 Through a visual inspection, we found that the axial outflow appears to be longer than 4~arcmin as reported by Becker et al. (2006) (i.e., the regions illustrated by R1 and R2 in Fig.~\ref{fig:images}), which was based only on the first three {\it XMM-Newton} observations.
 Apparently, the feature has an extension to region R3 and possibly farther out to R4. 
 Furthermore, besides the axial outflow, there appear to be two fainter bilateral features which have deviated from the proper motion direction (referred to as lateral outflows hereafter). 
 They are highlighted by the regions S1+S2 and N in Fig.~\ref{fig:images}. 

 To examine whether these features are genuine,  we compute their signal-to-noise ratios (S/N) from the unsmoothed raw images by following Li \& Ma (1983) with nearby source-free regions adopted for background estimation.
 For a conservative analysis, we removed the point-like sources detected by a source detection algorithm in the field by using XMMSAS task \texttt{edetect\_chain} for all subsequent analysis so as to avoid contamination by the background/foreground sources. The point sources were excised with circular regions with radii of 20 arcsec, which corresponds to an encircled energy fraction of $\sim80$\%.
 
 For the axial outflow, it is bright enough that we are able to examine its S/N ratios in individual observations. 
 The results are summarized in the upper panel of Table~\ref{tab:SNratio}.
 We consider a feature as genuine if the S/N ratio is larger than 3   in a single observation over the course of this work.
 Apart for the R2 region on 29 April 2004, which has a rather short effective exposure in the combined MOS1/2 data ($\sim10$~ks), both R1 and R2 regions of the axial outflow are set above our pre-defined detection threshold in all other observations.
 For R3, it is found to be above the threshold in half of the existing observations.
 For R4, its S/N is below 3 in most of the observations, except for the one on 14 April 2013.
We also compute the S/N ratios of R1, R2, R3, and R4 from the merged data as:\  $\sim27$, $\sim20$, $\sim13$, and $\sim8,$ respectively (see the lower panel of Table~\ref{tab:SNratio}).

 For the lateral outflows, since they appear to be fainter than the axial outflow, we computed their S/N ratios only in the merged data (cf., lower panel of Table~\ref{tab:SNratio}).
 The S/N ratios of N, S1, and S2 are found to be $\sim13$, $\sim16$, and $\sim16,$ respectively.

\begin{figure*}
        \begin{center}
        \includegraphics[width=1.5\columnwidth,trim={0cm 0cm 0cm 0cm},clip]{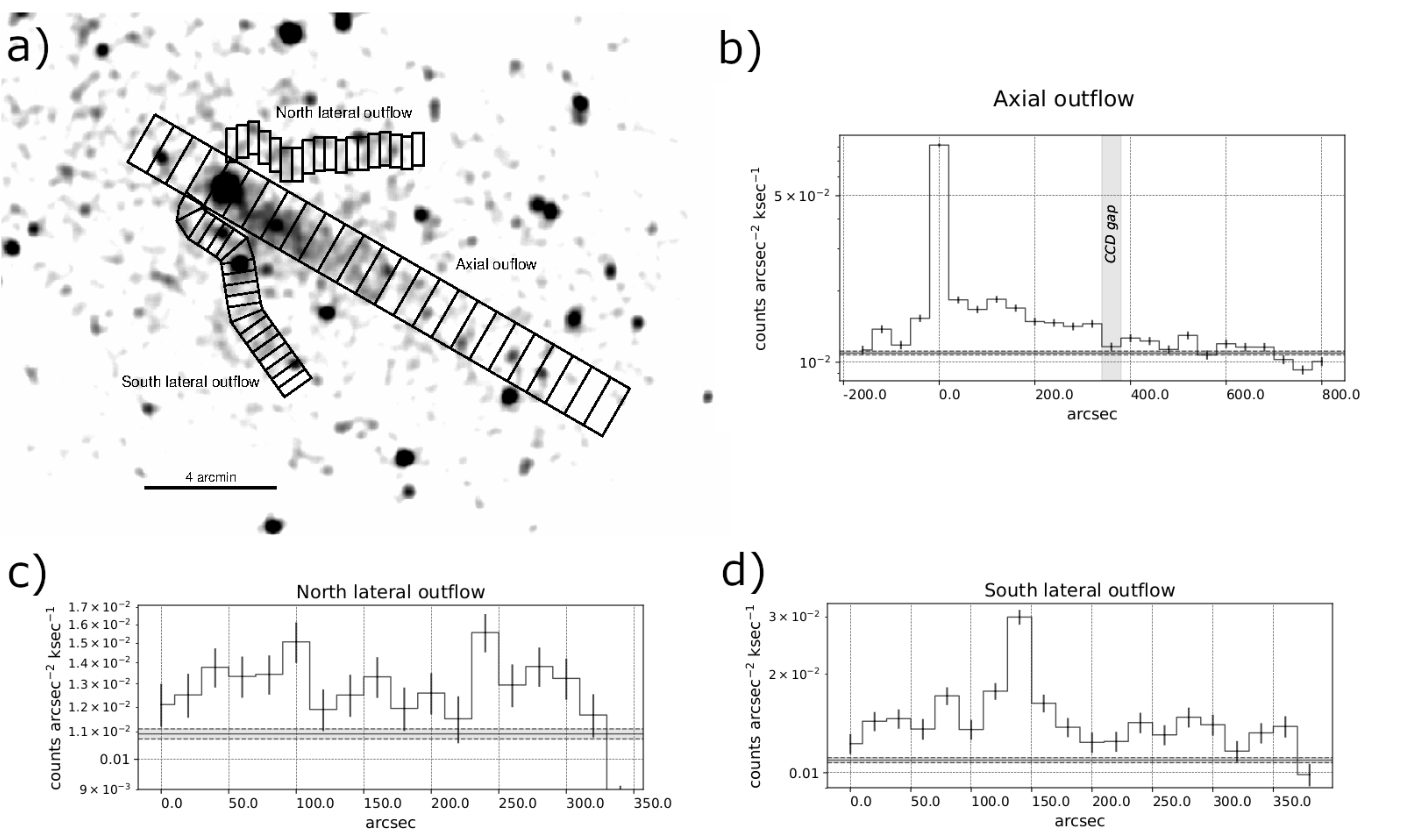}
        \end{center}
        \caption{(Panel a) Sampling regions for computing the brightness profiles of various spatial components. Brightness profiles of axial outflow (Panel b), north lateral outflow (Panel c) and south lateral outflow (Panel d) as estimated from the merged data. The black horizontal lines and the black dashed lines in Panels b, c and d indicate the background flux and its 1-${\sigma}$ error deviations, respectively.}
\label{fig:Briprf}
\end{figure*}

 To further quantify the size of each spatial component, we inspect their brightness profiles by using the unsmoothed raw images. 
 We first examined the axial outflow. Its profile is constructed by sampling in 25 consecutive boxes, each with a size of 40$\arcsec$$\times$100$\arcsec,$ as illustrated in Fig.~\ref{fig:Briprf}.
 We estimate the background levels by sampling six nearby source-free circular regions with the radius of $\sim74\arcsec$.
 In this work, we determine the size of the feature as the length from the pulsar position until it fades to a level that is consistent with that of the background.

 Since the axial outflow is the brightest component of the X-ray nebula associated with PSR~B1929+10, it is possible to examine if there is any spatial variability.
 By comparing its brightness profiles constructed from seven different observations, we cannot identify any significant variation in length.
 This prompts us to combine all the data so as to maximize the $S/N$, which allows us to place a better constraint on the size of the axial outflow. 
 The result is displayed in the upper-right panel in Fig.~\ref{fig:Briprf}. The black horizontal line and the black dashed line indicate the background flux and its 1-${\sigma}$ error deviations, respectively.
 It appears to fall to the background level at $\sim10$~arcmin behind the pulsar position, which corresponds to region R3 in Fig.~\ref{fig:images}.  Hence, we consider R3 to be genuine and to be an extension of the axial outflow from R1 and R2. 
 The discontinuation between R2 and R3 is due to the presence of a CCD gap in most of the adopted data. In view of the S/N  in R4, which is $<3$ in most observations, with no indication of any excess above the background level (as seen in Fig.~\ref{fig:Briprf}), we conclude that R4 was likely to have resulted from background fluctuation and thus, it will not be further considered in this work.

 For the lateral outflows, the photon statistics from individual observations are not sufficient for examining spatial variability.
 Therefore, we simply construct their brightness profiles from the merged data.
 The results are shown in the lower-left and lower-right panels in Fig.~\ref{fig:Briprf}.
 Both of them are found to have an extent of $\sim5$~arcmin.

\begin{table}
        \caption{($upper$ $panel$) Signal-to-noise ratios (S/N) ratios of each spatial component as estimated from the combined MOS1/2 images in individual observations. ($lower$ $panel$) S/N of different spatial components as estimated from the merged data.}
        \begin{center}
        \begin{tabular}{ccccc}
                \hline
                \hline
                Obs. Date   &    R1    &    R2    &    R3    &    R4    \\
                \hline
                11 Nov 2003 &   6.77   &  5.28   &   4.31   &  1.11   \\
                27 Apr 2004 &   6.09   &  5.02   &   1.48   &  1.99   \\
                29 Apr 2004 &   5.62   &  2.82   &   0.55   &  1.37   \\
                30 Mar 2013 &   7.27   &  4.23   &   2.71   &  0.16   \\
                30 Mar 2013 &   7.39   &  5.47   &   3.43   &  0.16   \\
                14 Apr 2013 &   11.52  &  9.98   &   5.71   &  3.85   \\
                14 Apr 2018 &   13.30  &  7.21   &   3.82   &  1.43   \\
                \hline
                \hline
        \label{tab:SNratio}
        \end{tabular}

        \begin{tabular}{cccc|ccc}
                \hline
                \hline
         \multicolumn{7}{c}{Merged Data}\\
                \hline
          \multicolumn{4}{c|}{Axial outflow}   &  \multicolumn{3}{c}{Lateral outflow} \\
          \hline
 R1   &    R2    &    R3    &    R4   &   N   &   S1   &   S2  \\
                \hline
27.02 &  20.38   &  12.56   &   8.00  & 13.18 & 15.62  & 16.02 \\
                \hline
                \hline
        \label{tab:SNmerged}
        \end{tabular}
        \end{center}
\end{table}

\subsection{Spectral analysis}
 We also performed a spatially resolved spectroscopy of the axial and lateral outflows.
 The source regions adopted for extracting the spectra are identical to those we used in the S/N calculations (see Fig.~\ref{fig:images}).
 Similarly, all the resolved point-like sources are removed before extracting the source spectra from the PWN. 

\renewcommand{\arraystretch}{1.3}
\begin{table*}
        \caption{Results of the spatially-resolved spectroscopy of PSR~B1929+10's X-ray nebula.}
        \begin{center}
        \begin{tabular}{c|cccccc}
                \hline
                \hline
\multirow{2}{*}{Component} & \multirow{2}{*}{Net count} & \multirow{2}{*}{$\chi^{2}$} & \multirow{2}{*}{dof} & \multirow{2}{*}{$\Gamma$} & Normalization & $F_{X}$ \\
              &      &                  &     &              &($10^{-5}$~photon/keV/cm$^{2}$/s at 1~keV)    & ($10^{-14}$ ergs cm$^{-2}$ s$^{-1}$) \\
                \hline
                \multicolumn{7}{c}{Axial outflow} \\
                \hline
R1     & 1168 & 52.66 & 55 & $2.64^{+0.13}_{-0.12}$ & $1.62\pm{0.09}$        & $7.86^{+0.52}_{-0.46}$\\
\hline
R2     &  809 & 64.18 & 57 & $2.28^{+0.18}_{-0.17}$ & $0.97\pm{0.09}$        & $4.86^{+0.33}_{-0.32}$\\
\hline
R3     &  427 & 32.04 & 48 & $1.80^{+0.29}_{-0.27}$ & $0.51^{+0.10}_{-0.09}$ & $3.27^{+0.44}_{-0.41}$\\
\hline
                \multicolumn{7}{c}{Lateral outflow} \\
                \hline
N      & 359  & 59.87 & 62 & $3.61^{+0.68}_{-0.53}$ & $0.63\pm{0.09}$ & $4.36^{+2.87}_{-1.35}$  \\
\hline
S1     & 550  & 73.84 & 63 & $2.45^{+0.25}_{-0.23}$ & $0.73\pm{0.08}$ & $3.54^{+0.39}_{-0.33}$ \\
\hline
S2     & 667  & 77.47 & 61 & $2.96^{+0.34}_{-0.30}$ & $1.10\pm{0.13}$ & $5.65^{+1.40}_{-0.89}$ \\
\hline
\hline
        \end{tabular}
        \end{center}
        \label{tab:FittingResults}
\end{table*}

 For the axial outflow, its spectra are sampled from regions R1, R2, and R3.
 For the south lateral outflow, we extract the spectra from regions S1 and S2.
 And for the north lateral outflow, as it is fainter, we extract its spectrum from a single region N instead. 
 Background spectra are extracted from the nearby source-free regions.
 After background-subtraction, the net counts of each spatial component collected from all observations are given in Column~2 of Table~\ref{tab:FittingResults}.
 All the response files are generated for each camera in individual observation with the aid of XMMSAS.

 All the spectral fittings in this work were performed with XSPEC (version 12.9.1m).
 We  adopted an absorbed power-law model for all the spatial components in the nebular emission.
  For the column absorption $N_{H}$, we adopted $N_{H}=2.40^{+0.36}_{-0.34}\times10^{21}$ cm${^{-2}}$ , as estimated from a fitting of the pulsar spectrum (see Hui \& Becker 2008).
 In order to better constrain the possible spectral variation among different nebular components, $N_{H}$ is fixed at the best-fit value inferred by Hui \& Becker (2008) throughout this work. 

 For each spatial component, we first examined the spectra obtained in different observations individually.
 We did not find any significant variability.
 The spectral parameters obtained from different observations were found to be consistent with each other within the tolerance of their statistical uncertainties. 
 Therefore, for each spatial component, we proceeded to fit the spectra obtained from all observations simultaneously with  a maximum likelihood approach so as to obtain a tight constraint on their emission nature. 
The results are summarized in Table~\ref{tab:FittingResults}. 

 For the axial outflow, we found that its X-ray spectrum is getting harder (i.e., the photon index $\Gamma$ is getting smaller) systematically along the direction away from the pulsar. This suggests evidence for spectral hardening, which may imply that a re-acceleration process might be present across the feature.  We examined the significance of variation by 
 fixing the photon index at $\Gamma=2.64$ (i.e., the best-fitted point estimate in R1) in R2 and R3 and the corresponding goodness-of-fit is degraded to namely $\chi^{2}=67.86$ (58 dof) and $\chi^{2}=38.33$ (49 dof) for R2 and R3, respectively.

\begin{figure}[h!]
        \begin{center}
        \includegraphics[width=\columnwidth,trim={0.5cm 0.3cm 0.7cm 0.5cm},clip]{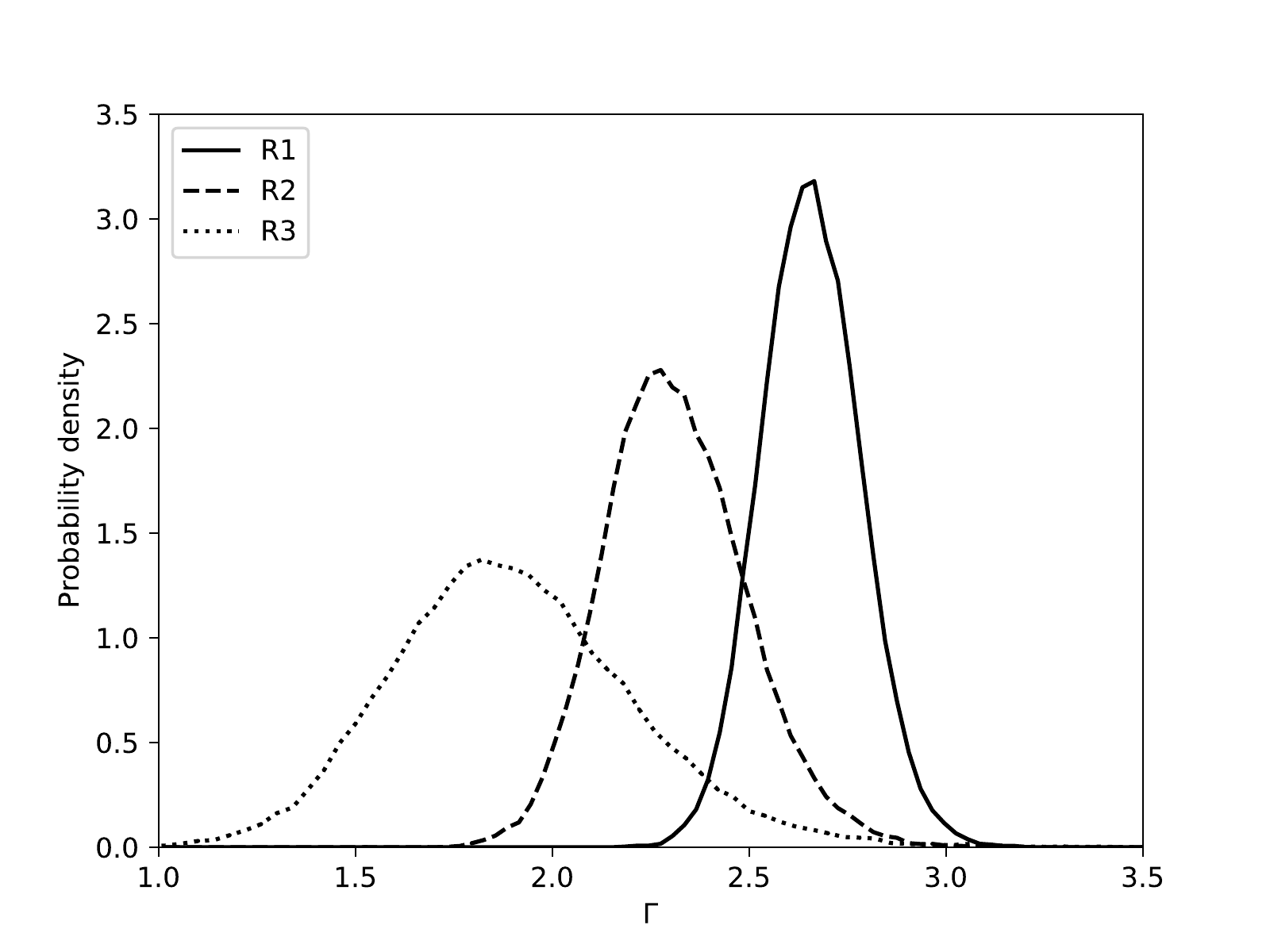}
        \includegraphics[width=\columnwidth,trim={0.5cm 0.3cm 0.7cm 0.5cm},clip]{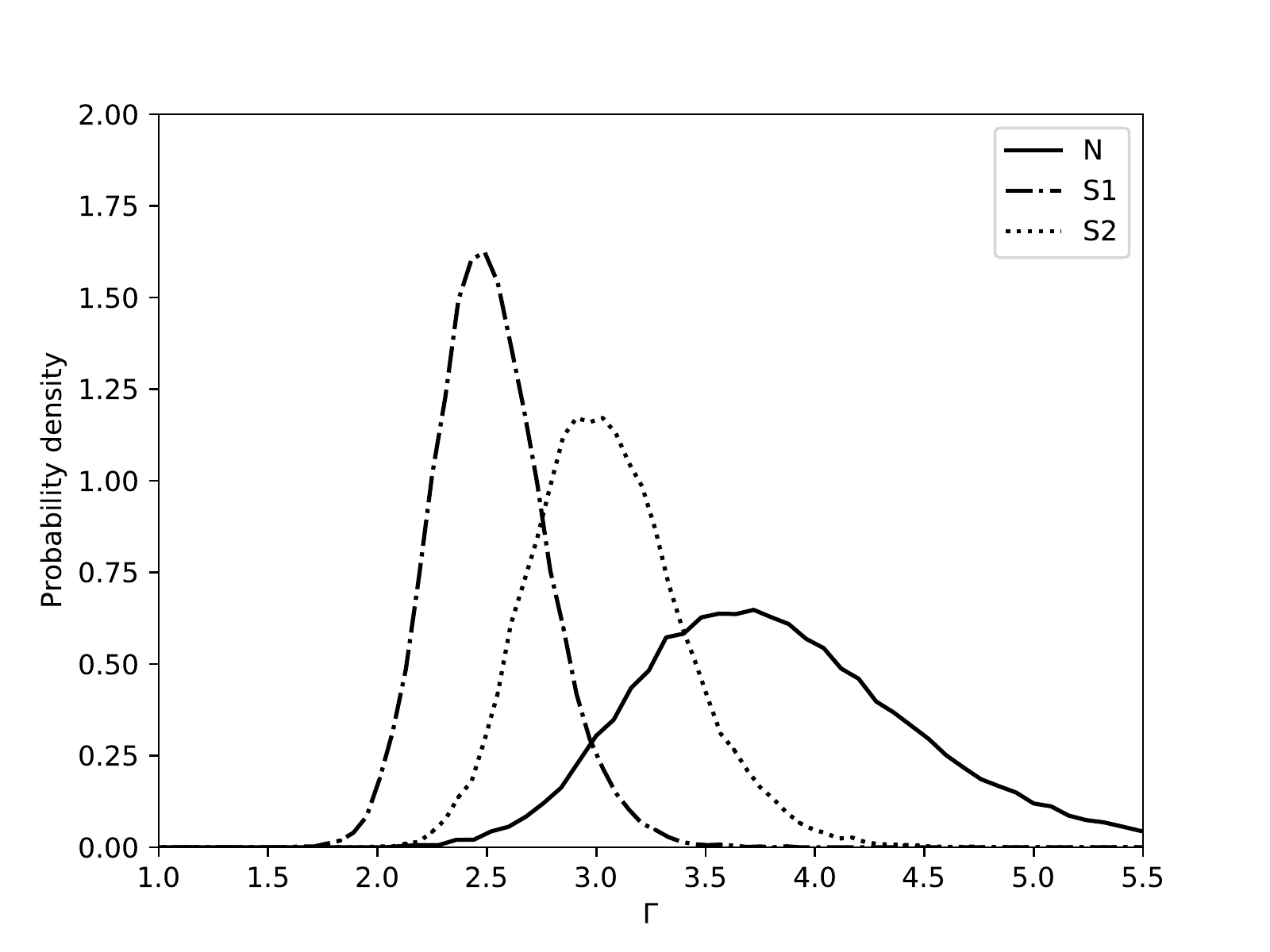}
        \caption{Posterior probability density of the photon indices of different regions of the axial outflow ($upper$ $panel$) and the lateral outflows ($lower$ $panel$).}
        \label{fig:MCMC}
        \end{center}
\end{figure}

 To further analyze this issue,  we did the Bayesian parameter estimation by assuming uniform priors for both the photon index and normalization. We used the 2D posterior probability density distribution for $\Gamma$ and the model normalization by Markov Chain Monte Carlo (MCMC). 
 We adopted the sampling algorithm devised by Goodman \& Weare (2010) with eight walkers, each taking 100000 steps, and we used 800000 samples to approximate the posterior probability density distribution. In the upper panel of Fig.~\ref{fig:MCMC}, we show the 1D marginalized posterior probability density distributions for the best-fit photon indices for R1, R2, and R3.
 The trend of spectral hardening along the axial outflow can be clearly seen 
in this figure.
 We have examined the robustness of our results by repeating the spectral fitting with a number of background spectra sampled from different regions.
 We found that the results from all these independent analyses are consistent, leading us to the conclusion that there is spectral hardening taking place along the axial outflow.

 On the contrary, the spectral fits for regions S1 and S2 of the south lateral outflow suggest spectral softening across the feature (Table~\ref{tab:FittingResults}, the lower panel of Fig.~\ref{fig:MCMC}).
 On the other hand, for the north lateral outflow (i.e., region N), its X-ray emission is found to be the softest ($\Gamma\sim3.6$) among all the identified nebular components.

 Adopting the best-fit spectral models and assuming a distance of $d=361$~pc, we estimated the absorption-corrected X-ray luminosity $L_{X}$ of each component in $0.3-10$~keV.
 The results are summarized in Table~\ref{tab:Luminosities}.
 Assuming they are powered by the rotation of the neutron star, we also computed the conversion efficiency $L_{X}/\dot{E}$ and the results are displayed as Column 3 in Table~\ref{tab:Luminosities}. 

\begin{table}
        \caption{X-ray luminosities and the conversion efficiencies in 0.3-10 keV of the different spatial components.}
        \begin{center}
        \begin{tabular}{c|c|c}
\hline
\hline
\multirow{2}{*}{Component} &     ${L_X}$     & \multirow{2}{*}{$L_X/\dot{E}$}\\
                           & (ergs s$^{-1}$) &              \\
\hline
\multicolumn{3}{c}{Axial outflow}\\
\hline
R1 & 1.23$^{+0.08}_{-0.07}\times10^{30}$ & 3.16$^{+0.21}_{-0.18}\times10^{-4}$\\
R2 & 7.62$^{+0.52}_{-0.50}\times10^{29}$ & 1.95$\pm{0.13}\times10^{-4}$\\
R3 & 5.13$^{+0.69}_{-0.64}\times10^{29}$ & 1.31$^{+0.18}_{-0.17}\times10^{-4}$\\
\hline
\multicolumn{3}{c}{Lateral outflow}\\
\hline
N  & 6.84$^{+4.50}_{-2.12}\times10^{29}$ & 1.75$^{+1.15}_{-0.54}\times10^{-4}$\\
S1 & 5.55$^{+0.61}_{-0.52}\times10^{29}$ & 1.42$^{+0.16}_{-0.13}\times10^{-4}$\\
S2 & 8.86$^{+2.19}_{-1.40}\times10^{29}$ & 2.27$^{+0.56}_{-0.36}\times10^{-4}$\\
\hline
\hline
        \end{tabular}
        \end{center}
        \label{tab:Luminosities}
\end{table}

\section{Discussion and summary}
 Here we present the results of the deep X-ray spectral imaging of the bow-shock PWN associated with PSR~B1929+10 by using a $\sim300$~ks effective exposure acquired by {\it XMM-Newton}.
 Besides the axial outflow behind the proper motion direction, which was discovered and investigated in the previous studies, we have identified two faint lateral outflows that extend laterally with respect to the proper motion (cf., Fig.~1 \& Fig.~2). 

 For the axial outflow, we found that it has an extent of $\sim8$~arcmin which is twice as the length deduced in the previous study, by using a much shorter exposure (Becker et al. 2006).
 Interestingly, we found the X-ray emission of the axial outflow is getting systematically harder  in the opposite direction of the pulsar's proper motion (see the upper panel of Figure~3).
 For the simplest synchrotron-emitting scenario that presumes the leptons are accelerated at the termination shock and advected into the backward direction, it is expected that the synchrotron spectrum should become softer at a distance that is further away from the pulsar.
 This assumption is contrary to our results. 

 Recently, our understanding of the nature of bow-shock PWN has been improved in the light of 3D relativistic magnetohydrodynamics (MHD) simulations (see Amato 2020 for an updated review).
 The numerical and analytical studies performed by Barkov et al. (2019) have shown that the magnetic stresses in the shocked wind have a dominant effect on the overall morphology and the dynamics of the bow-shock nebula.
 When the magnetic field and the outflow in the forward direction of the pulsar motion are turned backward near the head of the bow shock, they can partially contribute  to the axial outflow, together with the original tailward outflow.
 The magnetic field of the turned-back flow can have a different direction from that of the tailward outflow.
 And this can provide favorable locations for magnetic reconnection to take place in the axial outflow (see Fig.~11 in Barkov et al. 2019). 
 Through magnetic reconnections, magnetic energy can be converted into the kinetic energy of wind particles and this can lead to very efficient particle acceleration.
 This might provide an explanation for the observed spectral hardening in the axial outflow.

 For the lateral outflows, the orientations are similar to those have been observed in some other systems, such as Geminga (Hui et al. 2017; Posselt et al. 2017); PSR~J1509-5850 (Klingler et al. 2016).
 Such features can result from the polar outflows which are bent by the ram pressure (Hui et al. 2017).
 Another recent 3D MHD investigation performed by Olmi \& Bucciantini (2019) might shed light on the origin of these nebular components.
 By computing the trajectories of escaping electrons and positrons, Olmi \& Bucciantini (2019) have found that escaping flux can be highly asymmetry.
 Apart from the backward direction, a significant amount of 
wind particles can sometimes flow in the orthogonal directions with respect to the proper motion (see Figure 4 \& 5 in Olmi \& Bucciantini 2019).
The lateral outflows that we have seen may originate from these flow.

 Olmi \& Bucciantini (2019) further suggest that such asymmetric pulsar wind particle escape can be rather common. 
 However, the features that resemble the lateral outflow are typically faint.
 In our case of PSR~B1929+10, it requires an image with $\sim300$~ks exposure for revealing the lateral outflows. 
 Since these particle outflows can be related to the extended $\gamma-$ray halos, which is a growing population discovered by the \textit{High Altitude Water Cherenkov} (HAWC) observatory (Abeysekara et al. 2017) that can be a major contributor to the Galactic cosmic rays, deep X-ray imaging for the other bow-shock PWNe is therefore encouraged.

\begin{acknowledgements}
        SIK is supported by BK21 plus Chungnam National University, National Research Foundation of Korea grant 2016R1A5A1013277 and 2019R1F1A1062071; 
        CYH is supported by the National Research Foundation of Korea through grant 2016R1A5A1013277 and 2019R1F1A1062071; 
        JSL is suppored by the National Research Foundation of Korea grant funded by the Korean Government (NRF-2019H1A2A1077350Global Ph.D. Fellowship Program), 2016R1A5A1013277, 2019R1F1A1062071 and BK21 plus Chungnam National University;
        KMO is supported by National Research Foundation of Korea grant funded by the Korean Government (NRF-2019H1A2A1077058-Global Ph.D. Fellowship Program), 2016R1A5A1013277, 2019R1F1A1062071 and BK21 plus Chungnam National University;
        LCCL is supported by the National Research Foundation of Korea grant 2016R1A5A1013277; 
        JT is supported by the NSFC grants of China under 11573010, U1631103 and 11661161010.
\end{acknowledgements}

\begin{appendix}
\section{{\it XMM-Netwon} observations of PSR~B1929+10.}
\begin{table}[h!]
        \centering
        \caption{Exposures Of XMM-Newton Observations Of PSR B1929+10.}
        \label{tab:Exposures}
        \begin{tabular}{c c c c c} 
                \hline
                \hline
                Detector & \multirow{2}{*}{Filter} & Duration & \texttt{RATE} cut    & Exposure\\
                (EPIC)   &                         & (s)      & (Count/sec) &   (s) \\
                \hline
                \multicolumn{5}{c}{SEQ: 0718\_0113051301 2003 Nov 10}\\
                \hline
                  MOS1  & \multirow{2}{*}{Medium}  & 10670 & 1.3 & 8763.1   \\
                  MOS2  &                          & 10675 & 1.3 & 9537.8   \\
                \hline
                \multicolumn{5}{c}{SEQ: 0803\_0113051401 2004 Apr 27}\\
                \hline
                  MOS1  & \multirow{2}{*}{Medium}  & 21668 & 1.5 & 9445.0   \\
                  MOS2  &                          & 21673 & 1.5 & 10443.5   \\
                \hline
                \multicolumn{5}{c}{SEQ: 0804\_0113051501 2004 Apr 29}\\
                \hline
                  MOS1  & \multirow{2}{*}{Medium}  & 22669 & 3.2 & 4876.6   \\
                  MOS2  &                          & 22674 & 3.2 & 6080.7   \\
                \hline
                \multicolumn{5}{c}{SEQ: 2437\_0695200101 2013 Mar 30}\\
                \hline
                  MOS1  & \multirow{2}{*}{Medium}  & 21570 & 1.3 & 15249.6    \\
                  MOS2  &                          & 21624 & 1.5 & 14998.3   \\
                \hline
                \multicolumn{5}{c}{SEQ: 2437\_0695200201 2013 Mar 30}\\
                \hline
                  MOS1  & \multirow{2}{*}{Medium}  & 21570 & 1.3 & 14976.3   \\
                  MOS2  &                          & 21624 & 1.5 & 14033.2   \\
                \hline
                \multicolumn{5}{c}{SEQ: 2444\_0695200301 2013 Apr 14}\\
                \hline
                  MOS1  & \multirow{2}{*}{Medium}  & 61844 & 2.3 & 52024.4   \\
                  MOS2  &                          & 61901 & 3.2 & 52286.9   \\
                \hline
                \multicolumn{5}{c}{SEQ: 3360\_0804250401 2018 Apr 14}\\
                \hline
                  MOS1  & \multirow{2}{*}{Thin}    & 85342 & 1.6 & 41678.9   \\
                  MOS2  &                          & 85315 & 2.2 & 48821.6   \\
                \hline
                \hline
        \end{tabular}
\end{table}

\end{appendix}

\end{document}